\pdfoutput=1

\documentclass[11pt]{article}
\usepackage{tabularx}
\usepackage[preprint]{acl}
\usepackage{enumitem} 
\usepackage{times}
\usepackage{booktabs}
\usepackage{graphicx}
\usepackage{latexsym}
\usepackage{amsmath, amssymb}
\usepackage{algorithm}
\usepackage{algpseudocode}
\usepackage[T1]{fontenc}

\usepackage[utf8]{inputenc}

\usepackage{microtype}

\usepackage{inconsolata}

\usepackage{graphicx}

\usepackage{float}
\usepackage{multirow}
\usepackage{makecell}
\usepackage{adjustbox}
\usepackage{subfigure}

\usepackage{xcolor}
\usepackage{fvextra}
\usepackage{fancyvrb}
\usepackage[most]{tcolorbox}

\definecolor{redfg}{HTML}{D35B27}
\definecolor{redbg}{HTML}{FFF7F3}
\definecolor{yellowfg}{HTML}{FABB00}
\definecolor{yellowbg}{HTML}{FFF7E1}
\definecolor{bluefg}{HTML}{0C69DA}
\definecolor{bluebg}{HTML}{FBFCFE}

\title{ReSo: A Reward-driven Self-organizing LLM-based Multi-Agent System for Reasoning Tasks}

\author{
 \textbf{Heng Zhou\textsuperscript{1}\footnotemark[1]},
 \textbf{Hejia Geng\textsuperscript{2}\thanks{Equal contribution}},
 \textbf{Xiangyuan Xue\textsuperscript{1}},
 \textbf{Li Kang\textsuperscript{1}},
 \textbf{Yiran Qin\textsuperscript{1}}
\\
\textbf{Zhiyong Wang\textsuperscript{3}},
 \textbf{Zhenfei Yin\textsuperscript{3,4}\footnotemark[2]},
 \textbf{Lei Bai\textsuperscript{1}\thanks{Corresponding author,email:baisanshi@gmail.com, Primary Contact: Heng Zhou <hengzzzhou@gmail.com>}}, 
\\
\\
 \textsuperscript{1}Shanghai Artificial Intelligence Laboratory,
 \textsuperscript{2}Independent Researcher,
 \\
 \textsuperscript{3}The University of Sydney,
 \textsuperscript{4}Oxford University, 
 \\
}

\begin{document}
\maketitle
\begin{abstract}
Multi-agent systems have emerged as a promising approach for enhancing the reasoning capabilities of large language models in complex problem-solving. However, current MAS frameworks are limited by poor flexibility and scalability, with underdeveloped optimization strategies. To address these challenges, we propose ReSo, which integrates task graph generation with a reward-driven two-stage agent selection process. The core of ReSo is the proposed Collaborative Reward Model, which can provide fine-grained reward signals for MAS cooperation for optimization. We also introduce an automated data synthesis framework for generating MAS benchmarks, without human annotations. Experimentally, ReSo matches or outperforms existing methods. ReSo achieves \textbf{33.7\%} and \textbf{32.3\%} accuracy on Math-MAS and SciBench-MAS SciBench, while other methods completely fail. The code and data are available at \href{https://github.com/hengzzzhou/ReSo}{Reso}.
\end{abstract}

\section{Introduction}

Increasing inference time has emerged as a critical method to enhance the reasoning capabilities of large language models (LLMs)\cite{snell2024scalingllmtesttimecompute}. Two primary approaches have been explored: (1) optimizing a large reasoning model \cite{xu2025largereasoningmodelssurvey} by reinforcement learning and reward models during post-training, which could generate intermediate reasoning steps before answering  \cite{openai2024openaio1card, deepseekai2025deepseekr1incentivizingreasoningcapability} and (2) leveraging multi-agent system (MAS) collaboration to complete complex tasks that are difficult to solve by single inference \cite{han2024llm, guo2024largelanguagemodelbased, Wang_2024, tran2025multiagentcollaborationmechanismssurvey}.
\begin{figure}[ht]
    \includegraphics[width=1\linewidth]{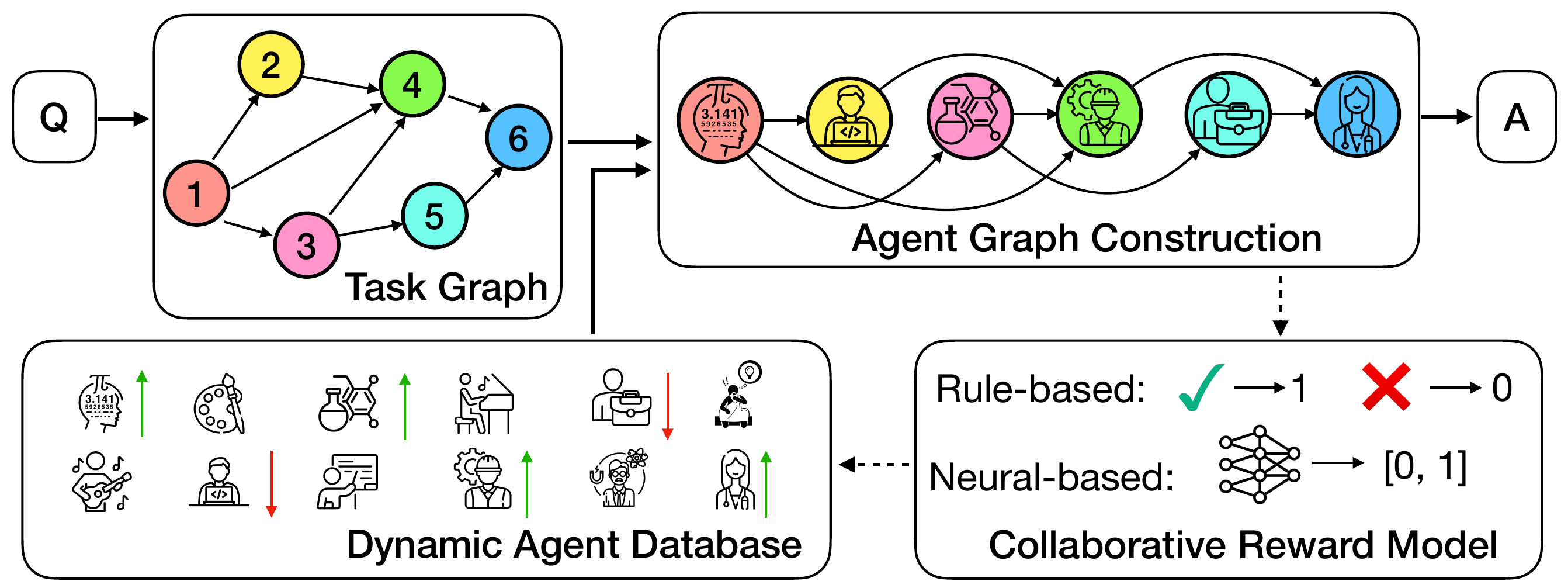}
    \label{fig:intro}
    \caption{Overview of ReSo pipeline. ReSo first decomposes the task into a DAG; and then constructs an agent graph by topological sorting. First, it searches for agent candidates for each subtask node from the dynamic agent database (DADB). Then it leverages the Collaborative Reward Model (CRM) to choose the best agent and update the agent estimation in DADB.}
\end{figure}
Compared to the success of inference time scaling on the single LLM, MAS faces multiple challenges. (1) Most are handcrafted, with limited scalability and adaptability. The lack of an effective agent self-organization mechanism hinders large-scale cooperation. (2) Most assume all agent abilities are fully known while assigning tasks, which is unrealistic for LLM-based agents. (3) Reward signals are restricted to missing, self-evaluation or outcome only, resulting in poorly defined optimization objectives. (4) Existing MASs lack mechanisms for dynamically optimizing agent networks, making it difficult to achieve data-driven improvements. To address these limitations, we ask: Can we design a self-organizing MAS to learn directly from data via reward signals without handcrafting?

To realize this potential, we propose ReSo, a reward-driven self-organizing MAS that integrates task graph generation and agent graph construction. The key innovation of our approach is the incorporation of fine-grained reward signals by the Collaborative Reward Model (CRM), which leads to dynamic optimization of agent collaboration. Different from existing MASs, our approach is both scalable and optimizable, achieving state-of-the-art performance on complex reasoning tasks.

While ReSo builds on prior work in agent selection and task decomposition, its principal contribution is the integrated formulation of these mechanisms within a self-organizing multi-agent reasoning framework. Our core insight is that individual agents exhibit heterogeneous expertise across different tasks and domains. During training, the CRM module evaluates each agent’s performance and records these scores in the DADB in \ref{dbda}. At inference time, ReSo decomposes a complex problem into subtasks and consults the DADB to dynamically assign each subtask to the agent best suited for it. This emergent, self-organizing process sets ReSo apart from traditional, linear pipeline architectures.
While extensive datasets exist for evaluating the reasoning capabilities of LLMs \cite{chang2023surveyevaluationlargelanguage, guo2023evaluatinglargelanguagemodels}, high-quality MAS evaluation benchmarks are scarce. Therefore, we propose an automatic data synthesis method to generate various MAS tasks by converting existing LLM benchmarks into complex collaboration problems. This method provides step-by-step reward signals without additional human annotations, enabling efficient and scalable MAS evaluation. Our contributions can be summarized as:
\begin{itemize}[%
    topsep=0pt, 
    partopsep=0pt,
    parsep=0pt, 
    itemsep=1pt   
]
    \item We first propose a Collaborative Reward Model, which can provide fine-grained reward signals for multi-agent collaboration.
    \item We present an automatic data synthesis method to generate arbitrarily complex MAS tasks from existing LLM benchmarks.
    \item We propose ReSo, the first scalable and optimizable self-organizing MAS framework. Experimental results demonstrate the superior performance of ReSo on challenging tasks.
\end{itemize}


\section{Related Work}

\subsection{Reward Guidance}

The reward model has become a critical component in enhancing the capabilities of LLMs through post-training \cite{wang2024reinforcementlearningenhancedllms}. By providing feedback on the quality of LLM outputs, RMs facilitate performance improvement, enabling models to generate more accurate and detailed responses. The concept of reward-guided learning was first introduced in InstructGPT \cite{ouyang2022traininglanguagemodelsfollow}, which uses human feedback to fine-tune LLMs, aligning their behavior with user intent. In addition to outcome-based supervision, process-based supervision has been shown to improve the reasoning process itself \cite{uesato2022solvingmathwordproblems}, enhancing not just the final answer but also the steps leading to it.

Building on this, \cite{lightman2023letsverifystepstep} introduced a process reward model (PRM) fine-tuned on PRM800K, which provides fine-grained and interpretable rewards for every reasoning step. Similarly, \cite{wang2024mathshepherdverifyreinforcellms} developed Math-Shepherd, an approach capable of autonomously generating process supervision data. Despite the advantages of neural-based reward models in terms of generalization, they also suffer from reward hacking \cite{gao2022scalinglawsrewardmodel, skalse2022definingcharacterizingrewardhacking}. To mitigate this, some recent approaches have employed rule-based rewards \cite{deepseekai2025deepseekr1incentivizingreasoningcapability} or fixed inference budgets \cite{muennighoff2025s1simpletesttimescaling}, which have also proven effective. Notably, DeepSeek-R1 \cite{deepseekai2025deepseekr1incentivizingreasoningcapability} incorporates both output accuracy and reasoning format evaluation, achieving the performance on par with OpenAI-O1 \cite{openai2024openaio1card, qin2024o1replicationjourneystrategic}. DeepSeek-R1 demonstrates that only using large-scale reinforcement learning based on rule-based reward during post-training can stimulate LLM’s excellent reasoning ability, without supervised fine-tuning.
 
\subsection{Multi-Agent System}
Recent advances in LLM-based MAS have raised expectations for their ability to tackle increasingly complex reasoning tasks \cite{han2024llm, guo2024largelanguagemodelbased, Wang_2024, tran2025multiagentcollaborationmechanismssurvey}.

Predefined cooperation in MAS relies on structured interactions and role assignments before collaboration. Early works focus on MAS infrastructure, including Camel, AutoGen, and AgentVerse \cite{li2023camelcommunicativeagentsmind, wu2023autogenenablingnextgenllm, chen2023agentversefacilitatingmultiagentcollaboration}. Some approaches adopt standard operating procedures for structured task decomposition, as seen in MetaGPT and ChatDev \cite{hong2024metagptmetaprogrammingmultiagent, qian2024chatdevcommunicativeagentssoftware, dong2024selfcollaborationcodegenerationchatgpt}. Fixed topologies are most adopted, such as hierarchical structures in MOA \cite{wang2024mixtureofagentsenhanceslargelanguage} and directed acyclic graphs in MacNet and MAGDI \cite{qian2024scalinglargelanguagemodelbasedmultiagentcollaboration, chen2024magdistructureddistillationmultiagent}. Predefined role interactions are also widely used such as debate \cite{du2023improvingfactualityreasoninglanguage}, criticism \cite{chen2024magicoremultiagentiterativecoarsetofine}, and certain math reasoning patterns \cite{gou2024toratoolintegratedreasoningagent,lei2024macmutilizingmultiagentcondition,xi2024enhancingllmreasoningcritique}. Predefined MASs exhibit several limitations including: (1) Scalability and adaptability being constrained by the imposition of rigid role assignments and fixed topological structures. (2) The unrealistic assumption that the agent’s abilities are fully known when assigning tasks, which is particularly problematic for LLM-based agents.

Optimizable cooperation in MAS aims to dynamically adapt interaction topology and agent roles. GPTSwarm \cite{zhuge2024languageagentsoptimizablegraphs} formulates MAS as optimizable computational graphs, refining node prompts and inter-agent connectivity via evolutionary algorithms. DyLAN \cite{liu2024dynamicllmpoweredagentnetwork} employs a layerwise feedforward agent network and a mutual rating mechanism to dynamically optimize MAS. G-Designer \cite{zhang2025gdesignerarchitectingmultiagentcommunication} utilizes variational graph auto-encoders to optimize MAS. Current optimizing approaches are highly underexplored. They often lack reliable, fine-grained reward signals for MAS collaboration, relying instead on outputs or self-generated reward mechanisms. Meanwhile, dynamic network optimization algorithms for MAS are also lacking.

\begin{figure*}[ht]
    \includegraphics[width=1\linewidth]{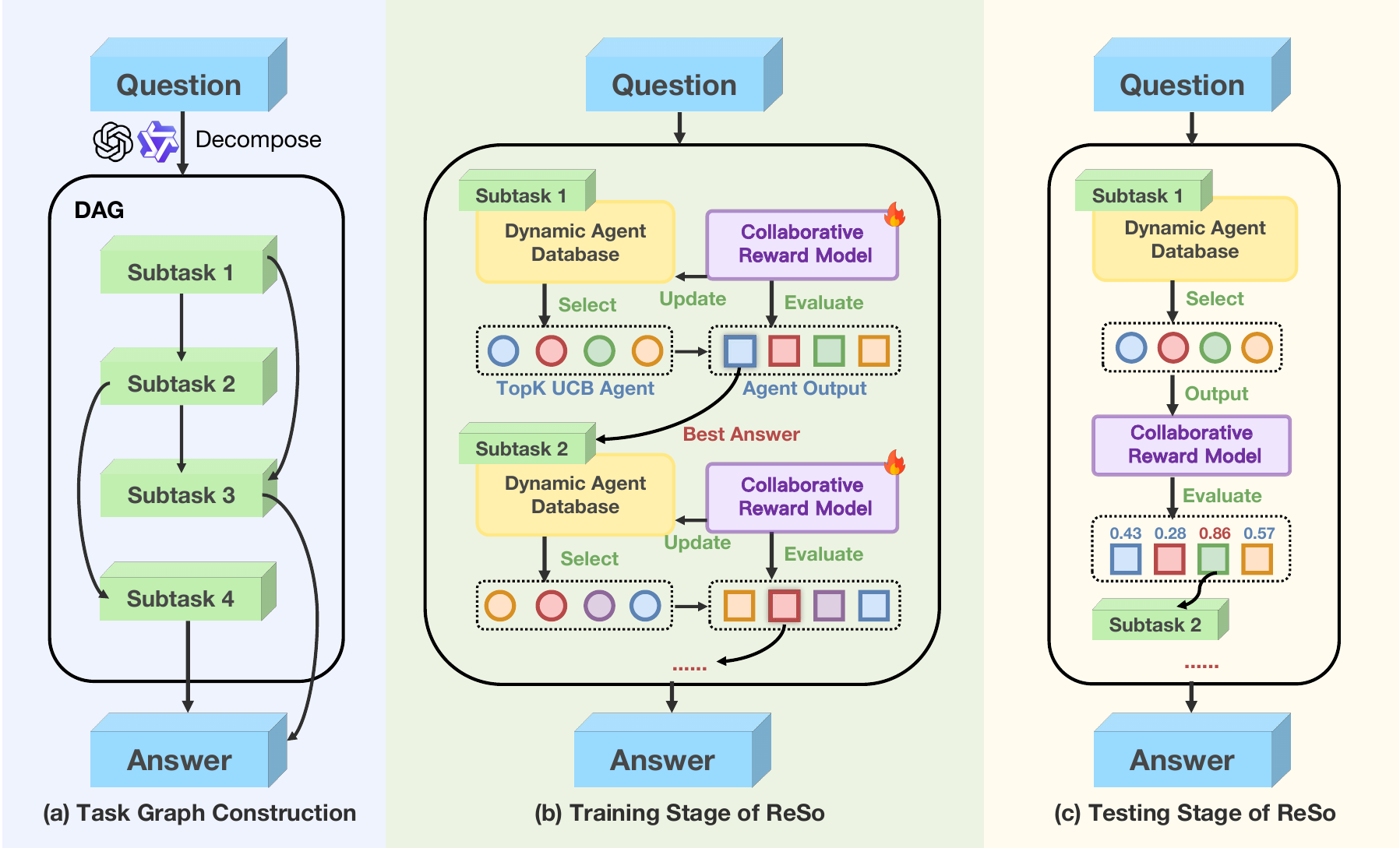}
    \label{fig:method}
    \caption{Illustration of our proposed ReSo. (a) We decompose the question into a subtask DAG. (b) The training of ReSo: we first use the UCB score to perform a coarse search in DADB and select top-k agents, then score the inference results using CRM, and update DADB by rewards. Repeat the above process for each node in DAG by topological order. (c) The testing of ReSo: we  select the best agent from DADB.}
\end{figure*}

\section{Methods}
To tackle the existing challenges in MAS research, we propose two core innovations: (1) ReSo, a reward-driven self-organizing MAS, which is capable of autonomously adapting to complex tasks and a flexible number of agent candidates, eliminating the need for handcrafted solutions. (2) Introduction of a Collaborative Reward Model (CRM), specifically tailored to optimize MAS performance. CRM can deliver fine-grained reward signals on multi-agent collaboration, enabling data-driven MAS performance optimization.

\subsection{Problem Formulation}
We define a MAS algorithm \( f_{MAS} \) as a function that, given a natural language question \( Q \), generates a graph-structured task decomposition, solves each subtask, and produces a final answer:
\begin{equation}
    f_{MAS}(Q) \rightarrow 
    \bigl(
    G = (V, E), \; A_V, \; A_Q
    \bigr)
\end{equation}

Here, \( {G} = ({V}, {E}) \) represents the task decomposition graph, which is structured as a directed acyclic graph (DAG). The set of nodes \( {V} = \{v_1, v_2, \dots, v_n\} \) corresponds to the subtasks derived from \( Q \), while the edges \( {E} \subseteq {V} \times {V} \) define the dependencies between these subtasks. The system produces subtask answers \( {A_V} = \{a_{v_1}, a_{v_2}, \dots, a_{v_n}\} \) and ultimately derives the final answer \( {A_Q} \). To achieve this, we decompose \( f_{MAS} \) into two sub-algorithms:
\begin{equation}
    f_{MAS}(Q) = f_{agent} \circ f_{task}(Q)
\end{equation}

 \( f_{task} \) is responsible for constructing the task decomposition graph from the input question, ensuring a structured breakdown of the problem into subtasks and dependencies. \( f_{agent} \) dynamically selects and assigns appropriate agents to solve the identified subtasks. This modular design enables independent optimization of each component, allowing for greater flexibility and scalability. 

For the MAS-generated answer \( A_Q \) to be considered correct, the following conditions must be satisfied: (1) All subtask answers must be correct. (2) All directed edges must correctly enforce the dependency relationships among subtasks. (3) The final output \( A_Q \) must be correct.  

\subsection{Task Graph Construction}
In the proposed method, \(f_{task}\) first transforms the question \(Q\) into a directed acyclic task graph \(G\):
\begin{equation}
    f_{task}: Q \;\;\rightarrow\;\; G = (V, E)
\end{equation}
where \(G\) represents the decomposition of the original task \(Q\). Each node \(v_i \in V\) is a natural language subtask, and each directed edge \((v_i \rightarrow v_j) \in E\) indicates that the subtask \(v_j\) depends on the successful completion of \(v_i\). 

In practice, we perform supervised fine-tuning (SFT) on an LLM to perform this step of task decomposition. Using our synthetic data, we explicitly require the LLM to decompose \(Q\) into logical sub-problems, specify their execution order and dependencies, and output in a format of DAG.

\subsection{Two-Stage Agent Search}
Once the task graph is obtained, we need to assign each subtask to the most appropriate agent. We denote this agent assignment procedure as \(f_{agent}\). Conceptually, \(f_{agent}\) classifies each node in the task graph according to the most suitable agent from a large agent pool $\mathcal{A}$, constructing an \textit{agent graph} that maps each node to one or more selected agents.
\begin{equation} 
f_{agent}: v_i \in V \;\;\rightarrow\;\; a_i \in \mathcal{A} \end{equation}
Since $\mathcal{A}$ can contain a large number of agents, we first introduce the concept of Dynamic Agent Database. Then we decompose the agent graph construction on every subtask into two search algorithms from coarse to fine-grained: first, select a subset of candidates from DADB then utilize the reward model to evaluate and select the best agent.

\subsubsection{Dynamic Agent Database}  
\label{dbda}
To increase MAS's scalability and flexibility, we propose the Dynamic Agent Database (DADB), denoted as $\mathcal{A}$, which enables adaptive agent selection by maintaining both \textbf{static} and \textbf{dynamic} agent profiles. For each agent $a_i \in \mathcal{A}$, its static profile includes the base model, role settings, initial prompt, long-term memory, and tools. The dynamic profile, continuously updated via the reward model, tracks the agent’s average reward \(R(a_i)\), computational cost \(C(a_i)\), and task count \(n(a_i)\). Initially, agents have only static attributes, while training iteratively refines their evaluations by the process reward model, optimizing future selection.

Given an input task \(v_j\), the DADB assigns a preliminary quality score \(Q(a_i, v_j)\) to each agent \(a_i\), balancing task-agent similarity, historical performance, and computational costs:
\begin{equation}
    Q(a_i,v_j) \;=\; \mathrm{sim}(a_i,v_j)\cdot \mathrm{perform}(a_i)
\end{equation} where \(\mathrm{sim}(a_i,v_j)\) represents the similarity between the subtask’s target profile and the agent’s static profile. In practice, we employ a Heaviside function which ensures that only agents exceeding a predefined similarity threshold \(V_{th}\) are considered:
$ \mathrm{sim}(a_i,v_j)\; = \; H[ \langle\mathbf{q_i},\mathbf{a_i}\rangle - V_{th}]
$ where $\mathbf{q_i}, \mathbf{a_i}$ are text embedding of subquestion and the agent static profile. The \(\mathrm{perform}(a_i)\) term is given by $\mathrm{perform}(a_i) \;=\; R(a_i) \;-\; \beta C(a_i)$, where \(\beta\) controls the trade-off between the agent's historical performance and cost.

\subsubsection{Coarse Agent Search by UCB}
Given a DADB $\mathcal{A}$ and a subtask $v_j$, our first objective is to retrieve a promising subset of \(k\) candidate agents. To take advantage of the known information in DADB, also to explore unused agents, we adopt an Upper Confidence Bound value:
\begin{equation}
    \text{UCB}(a_i,q_j) \;=\; Q(a_i,q_j) \;+\; c \sqrt{\frac{N}{n(a_i) + \varepsilon}} 
\end{equation}
where $N$ is the total number of agent selections and $n(a_i)$ the number of times agent $i$ is selected, $\varepsilon \ll1$. \(c\) is a constant controlling the exploration-exploitation trade-off. Agents with higher UCB scores are more likely to be selected, helping the MAS to explore potentially under-utilized agents. For each subtask $q_i$, we sort agents by their $\text{UCB}(a_i, q_j)$ and choose the top \(k\) agents as the candidate set \({A}_{\text{cand}} = \{\,a_1, a_2,\dots,a_k\}\).

\subsubsection{Fine-grained Agent Evaluation by CRM}
\label{sec:crm}
Once the candidate agents \(\mathcal{A}_{\text{cand}}\) are selected, we evaluate their performance on the current subtask \( v_j \) using a Collaborative Reward Model (CRM). This evaluation process is straightforward: each candidate agent \( a_i \) generates an answer to the subtask \( v_j \): \( a_i(v_j) \), and then we assess the quality of that answer based on a reward signal:
\begin{equation}
    r(a_i, v_j) \;=\; \text{RewardModel}\Bigl(a_i, v_j, a_i(v_j)\Bigr)
\end{equation}
where \(\text{RewardModel}\) evaluates the quality of the solution based on the given agent’s profile, subtask, and previous reasoning process. After evaluating the agents, we assign the agent with the highest reward, \( a_j^* \), to the subtask node \( v_j \), which means \( a_j^* \)’s solution is used as \( v_j \)'s answer. This process is repeated for each subtask on the graph.

The reward \( r(a_i, v_j) \) is computed using the CRM, which can be either rule-based (e.g., binary correctness: 0 for incorrect, 1 for correct) or neural-based (providing a score between 0 and 1 for quality). The reward model evaluates how well the agent’s response aligns with the expected outcome, factoring in both the solution's correctness and its collaboration within the MAS.

\subsection{Training and Inference Stage}
Our multi-agent system can operate in two modes: training and testing. During \textbf{training}, we leverage a high-quality reward $r(a_i, v_j)$ available for evaluating the correctness of every step of MAS. Upon receiving \(r(a_i, v_j)\) for each candidate agent, we update that agent's dynamic profile in DADB. For instance, we may maintain a running average of rewards:
\begin{equation}
    R(a_i) \;\leftarrow\; \frac{n(a_i)\cdot R(a_i) + r(a_i, v_j)}{n(a_i)+1}
\end{equation}
similar for updating $cost c(a_i, v_j)$. By iteratively learning from data, the DADB can dynamically update agent evaluations based on historical reward, facilitating adaptive agent selection and improving both efficiency and performance. During \textbf{testing}, the reward model is no longer required. Instead, we leverage the learned DADB to select the best agent candidates and the best answer to each subtask.

\subsection{The Perspective of MCTS}The task graph, after topological sorting, forms a decision tree where each node represents a subtask and the edges denote dependencies. At each level, we use UCB to prune the tree and select a subset of promising agents, then simulate each agent and evaluate their performance using the CRM. The resulting reward updates the agent’s dynamic profile, refining the selection strategy. The MAS construction is essentially finding the optimal path from the root to the leaves, maximizing the UCB reward for the best performance.

Consider there are \( N \) agents and a task requiring \( D \) agents to collaborate. Assume that the average inference cost is $c$ and the matching cost in DADB is $s \ll c$ per agent. A brute-force search has a complexity of \( O(c \cdot N^D) \), which becomes infeasible as \( D \) and \( D \) grow. In contrast, our self-organizing strategy, selecting top$k$ per step, reduces the cost to \( O((s \cdot N + N \log N + k \cdot c) \cdot D) \), offering a near-linear scaling with \( N \) and \( D \), making the approach highly scalable for large $N$ and $D$.

\section{Data Synthesis}
\label{sec:data_synthesis}

A key challenge in MAS is the lack of structured datasets for evaluating and training agent collaboration. To address this, we propose an automated framework that converts existing LLM datasets into structured, multi-step MAS tasks, enabling fine-grained evaluation without human annotations.

\paragraph{Random DAG Generation}We begin by generating a DAG, \( G = (V, E) \). Each node \( v_i \in V \) will be filled with a subtask \((q_i, a_i)\), where \( q_i \) is the textual description of the task, and \( a_i \) is its numerical answer. The subtasks are sampled from the existing LLM benchmarks. The edges \( E \) will encode dependency constraints between subtasks, ensuring that the solution to one subtask is required as an input for another, modeling the sequential reasoning process of multi-agent collaboration.

\paragraph{Subtask Selection and Filling}  To populate the nodes of \( G \), we construct a master pool of candidate subtasks, denoted as \( \mathcal{P} \). Each candidate subtask \( p_i \in \mathcal{P} \) consists of a textual problem description \( s_i \), and a numerical answer \( a_i \). After obtaining $\mathcal{P}$, we randomly sample from it and fill one question per node into the generated DAG. Candidate subtasks should have clear numerical or option answers, such as SciBench \cite{wang2024scibenchevaluatingcollegelevelscientific}, Math \cite{hendrycks2021measuringmathematicalproblemsolving}, GPQA \cite{rein2023gpqagraduatelevelgoogleproofqa}, etc. To ensure that the problem is computationally feasible for later dependency construction, we extract a numerical constant \( c_i \in \mathbb{R} \) from the problem text. If the extracted constant is valid, the subtask is retained in \( \mathcal{P} \); otherwise, it is discarded. This ensures that only problems with well-defined numerical attributes are incorporated.

\paragraph{Dependency Edge Construction}After all nodes are populated, we generate natural language dependency descriptions for edges. Each edge \( (v_j \to v_k) \) should represent a relationship which connects previous subtask \( v_j \)'s answer \( a_j \), with subsequent subtask \( v_k \)'s question parameter \( c_k \). For each edge, we generate a textual description $e_{jk}$, such as ``in this question, $c_k$ = previous answer + 3.'' Formally, it is an algorithm that constructs a string from two numbers: $e_{ij} = f(a_j, c_k)$. $f$ can be implemented using elementary arithmetic and text templates, ensuring that no answers or parameters in the original subtask need to be manually modified. Once the DAG is fully constructed, we refine node descriptions by removing any explicitly given numerical constants $\{c_i\}$ that are now dependent on the results of prior nodes. Finally, an entire graph described in natural language is a piece of synthetic data.
 
The proposed data synthesis framework generates structured, multi-step reasoning tasks with adjustable sizes, ensuring diverse and scalable problem structures. The synthesized dataset supports both training and testing, enabling fine-grained evaluation without human annotations.

\section{Experiments}
\label{sec:experiment}

In \ref{subsec:exp_datamodel}, we first use public datasets to create complex MAS benchmarks and fine-tune ReSo’s task decomposition and collaborative reward models. All code, datasets, and models are publicly available. In \ref{subsec:main_results}, we train and evaluate ReSo on both public and synthetic datasets. \ref{subsec:ablation} presents ablation studies on task decomposition, agent selection, and reward guidance mechanisms.

\subsection{Data Synthesis and Model Fine-tuning}
\label{subsec:exp_datamodel}

\subsubsection{Data Synthesis}

MATH~\cite{hendrycks2021measuringmathematicalproblemsolving} consists of problems from diverse mathematical domains, while SciBench~\cite{wang2024scibenchevaluatingcollegelevelscientific} includes scientific reasoning tasks spanning physics, chemistry, and mathematics. Using these datasets, we apply the synthetic data generation method outlined in \ref{sec:data_synthesis} to create two datasets: one for single LLM fine-tuning and another for benchmarking. Difficulty is categorized by the number of subtasks—Easy (3), Medium (5), and Hard (7).

\paragraph{Fine-tuning data} \label{para:finetunedata} For fine-tuning task decomposition LLM, we generate 14,500 questions and answers from the MATH training set, with numbers of subtasks ranging from 2 to 6. For fine-tuning the neural-based CRM, we generate 5,000 questions from the same set, with 5 subtasks per question.

\subsubsection{Model Fine-tuning}
\paragraph{Task Decomposition Model Training} To ensure high-quality task composition, we fine-tune a specialized model for task decomposition based on Qwen2.5-7B-Instruct. We use 14500 dialogues on task decomposition as described in \ref{para:finetunedata}, and fine-tune the model under a batch size of 128 and a learning rate of 1e-4 for 3 epochs. The fine-tuned model can reliably produce task decomposition in a structured format.

\paragraph{CRM Training} The proposed CRM is fine-tuned based on Qwen2.5-Math-PRM-7B~\cite{zhang2025lessons}, which can provide effective process reward signals on MAS collaborative reasoning tasks. We use 5000 samples of sub-tasks with their answers as described in \ref{para:finetunedata}. We follow a simplified training scheme of PRMs, where the model should only perform binary classification on the special token at the end of the answer. The model is trained with a batch size of 128 and a learning rate of 1e-4 for 5 epochs. The fine-tuned model can output the probability of the answer being correct, which is then taken as the collaborative reward signal.
\begin{table*}[ht]
    \renewcommand{\arraystretch}{0.9}
    \centering
    \begin{tabular}{lcccccccc}
    \toprule
    \multirow{2}{*}{\textbf{Method}} & \multicolumn{4}{c}{\textbf{Math-MAS}} & \multicolumn{4}{c}{\textbf{SciBench-MAS}} \\
    \cmidrule(lr){2-5} \cmidrule(lr){6-9}
    & Easy & Medium & Hard & \textit{Tokens} & Easy & Medium & Hard &  \textit{Tokens} \\
    \midrule
    GPT-4o & 27.5 & 9.0 & 0.0 & 2.2k & 39.3 & 12.5 & 1.6 & 2.1k \\
    Gemini-2.0-Flash & \underline{69.2} & \underline{24.7} & 9.0 & 3.0k & \underline{64.5} & \underline{33.8} & 9.7 & 2.5k \\
    Claude-3.5-Sonnet & 12.1 & 0.0 & 0.0 & 1.0k & 22.4 & 6.2 & 3.2 & 1.4k \\
    Qwen2.5-Max & 44.0 & 13.5 & 4.5 & 2.9k & 55.1 & 30.0 & 4.8 & 2.8k \\
    DeepSeek-V3 & 52.7 & \underline{24.7} & 12.4 & 2.2k & 52.3 & 31.3 & \underline{12.9} & 2.3k \\
    \midrule
    MetaGPT & 30.8 & 12.4 & 2.2 & 16.1k & 48.6 & 2.5 & 0.0 & 14.6k \\
    DyLAN & 40.7 & 9.0 & 0.0 & 64.1k & 48.6 & 2.5 & 0.0 & 77.8k \\
    GPTSwarm & 35.2 & 5.6 & 4.5 & 14.9k & 31.8 & 6.3 & 1.6 & 18.2k \\
    GDesigner & 14.2 & 5.6 & 0.0 & 16.9k & 24.3 & 12.5 & 0.0 & 19.0k \\
    \textbf{ReSo (ours)} & \textbf{79.1} & \textbf{56.2} & \textbf{33.7} & \textbf{14.6k} 
                         & \textbf{67.3} & \textbf{51.3} & \textbf{32.3} & \textbf{20.7k} \\
    \bottomrule
    \end{tabular}
    \caption{Accuracy and average token usage on Math-MAS and SciBench-MAS. Bold and underlined represent optimal and suboptimal results, respectively.
    \textit{Tokens} denotes the average number of tokens consumed per task.}
    \label{tab:model_performance}
\end{table*}

\paragraph{MAS Benchmarks}\label{para:masdata} We select 201 questions from SciBench as the sub-question data pool and synthesized complex data using the method in \ref{sec:data_synthesis}. This forms the SciBench-MAS dataset, comprising 200 easy-level training questions and 247 testing questions (107 easy, 80 medium, 62 hard). For MATH~\cite{hendrycks2021measuringmathematicalproblemsolving}, 348 level-5 questions are selected, from which we generate the Math-MAS dataset, consisting of 269 test questions for ReSo (91 easy, 89 medium, 89 hard).


\subsection{Main Results of ReSo}
\label{subsec:main_results}

\paragraph{Models and MASs}
We compare ReSo with state-of-the-art LLM and MAS methods. Our single-LLM baselines include GPT-4o~\cite{openai2024gpt4ocard}, Gemini-2.0-Flash~\cite{team2024gemini}, Claude-3.5-Sonnet~\cite{anthropic2024claude}, Qwen2.5-Max~\cite{yang2024qwen2}, DeepSeek-V3~\cite{liu2024deepseek}. For ReSo, we build an agent database that includes these base models, extended to 63 agents with different prompts. For MAS, we evaluate MetaGPT~\cite{hong2024metagptmetaprogrammingmultiagent}, DyLAN~\cite{liu2024dynamicllmpoweredagentnetwork}, GPTSwarm~\cite{zhuge2024languageagentsoptimizablegraphs}, GDesigner~\cite{zhang2025gdesignerarchitectingmultiagentcommunication}. All MAS baselines use GPT-4o as the backbone.

\paragraph{Comparisons with LLMs} As shown in Table~\ref{tab:model_performance}, most single-model agents exhibit a sharp decrease in accuracy as the difficulty increases. At the hard difficulty level, their accuracy approaches zero, suggesting that single LLMs struggle with compositional reasoning. In particular, we show the results of these single LLMs on single Math and Scibench datasets in Appendix~\ref{app:single_llm} , with accuracy rates of 80\%-90\%. This means that a single LLM can successfully solve a single sub-problem in the dataset, but its generalization ability for combined complex problems is very limited.

\paragraph{Comparisons with MASs} Notably, ReSo outperforms other approaches in both the Math-MAS and SciBench-MAS datasets. At the hard difficulty level, ReSo reaches an accuracy of \textbf{33.7\%} on Math-MAS and \textbf{32.3\%} on SciBench-MAS, while other MAS methods almost completely fail. 

\paragraph{Results on Standard Benchmarks} 
Our approach demonstrates robust performance not only on complex task datasets but also on widely adopted benchmarks. Table~\ref{tab:existing_benchmark} summarizes the comparative accuracy, where ReSo consistently achieves the highest scores across all tasks. These results attest to ReSo’s strong generalization capabilities and its effectiveness in mathematical and scientific reasoning, as well as related domains.

\begin{table}[ht]
  \centering
  \caption{Comparison of accuracy (\%) on standard benchmarks.}
  \resizebox{.48\textwidth}{!}{
    \begin{tabular}{lcccc}
      \toprule
      \textbf{Method}     & \textbf{GSM8K} & \textbf{GPQA} & \textbf{HumanEval} & \textbf{MMLU} \\
      \midrule
      DyLAN               & 88.16          & 49.55         & 89.70              & 80.16         \\
      GDesigner           & 95.07          & 53.57         & 89.90              & 84.50         \\
      GPTSwarm            & 89.74          & 52.23         & 88.49              & 83.98         \\
      \textbf{ReSo (ours)}& \textbf{95.70} & \textbf{55.80} & \textbf{92.00}     & \textbf{88.70}\\
      \bottomrule
    \end{tabular}
  }
  \label{tab:existing_benchmark}
\end{table}

\begin{figure*}[ht]
	\centering
    \subfigure[Task Decomposition]{
        \includegraphics[width=0.31\linewidth]{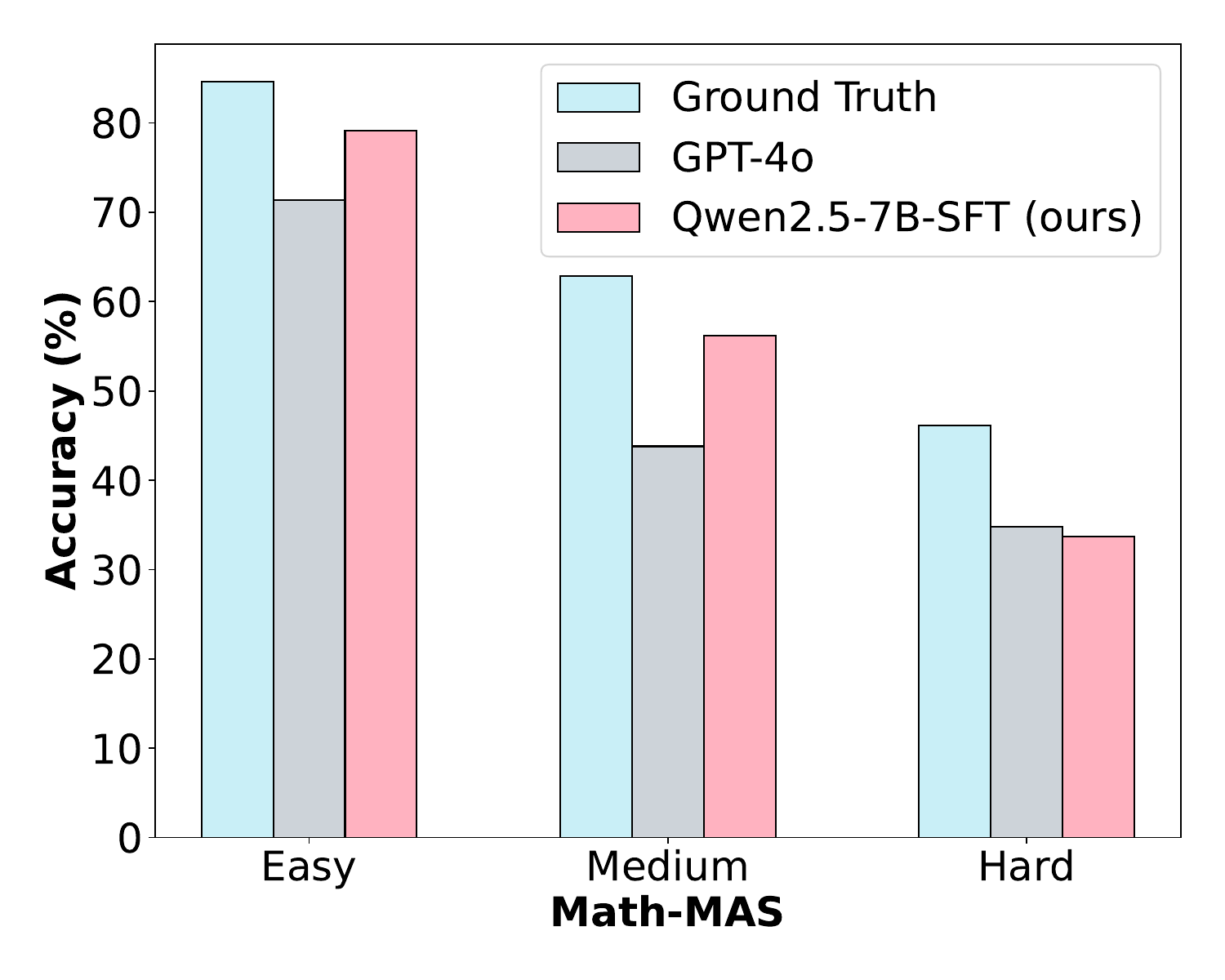}
        \label{fig:ablation_decomposition}
    }
    \subfigure[Agent Selection]{
        \includegraphics[width=0.31\linewidth]{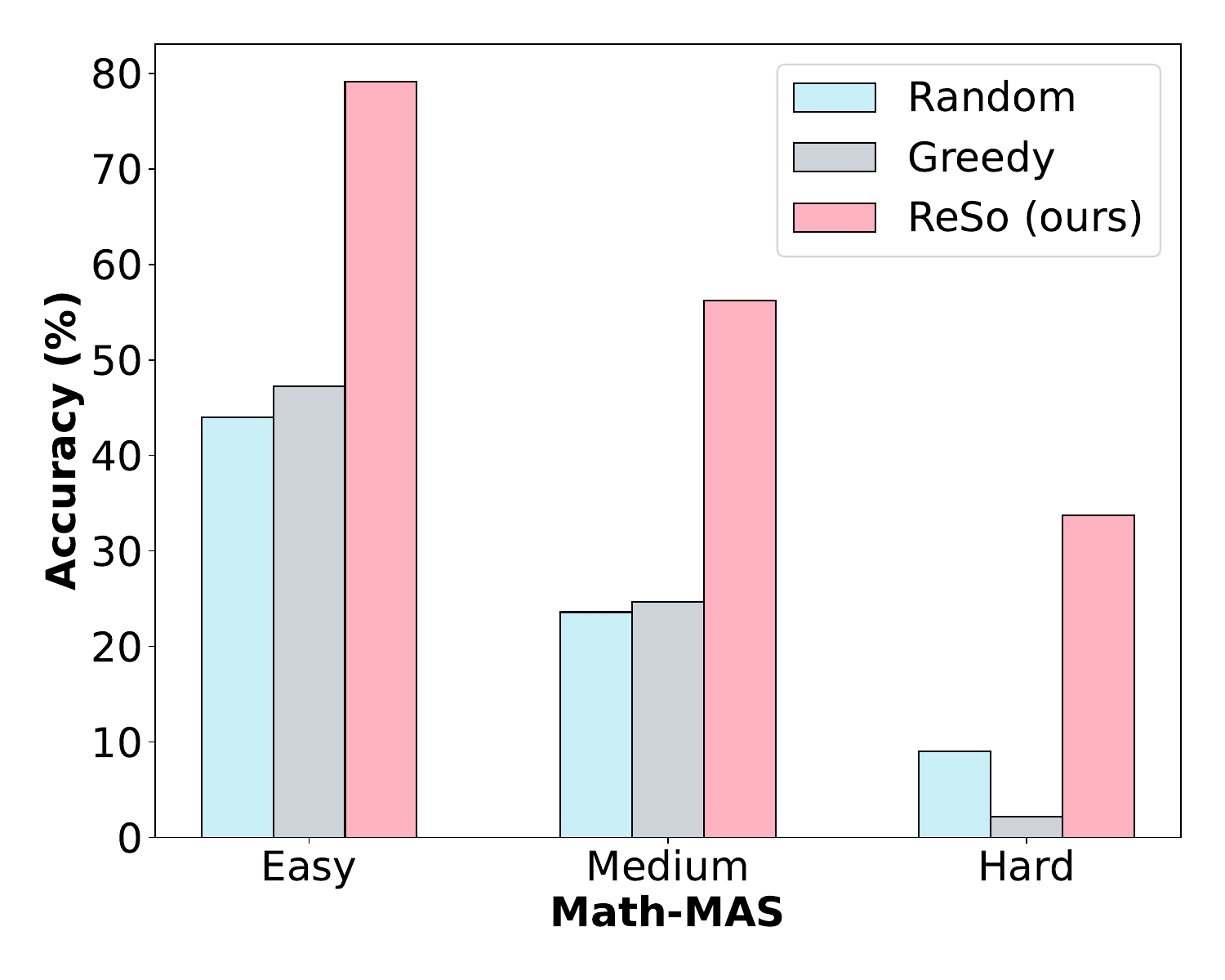}
        \label{fig:ablation_selection}
    }
    \subfigure[Reward Signal]{
        \includegraphics[width=0.31\linewidth]{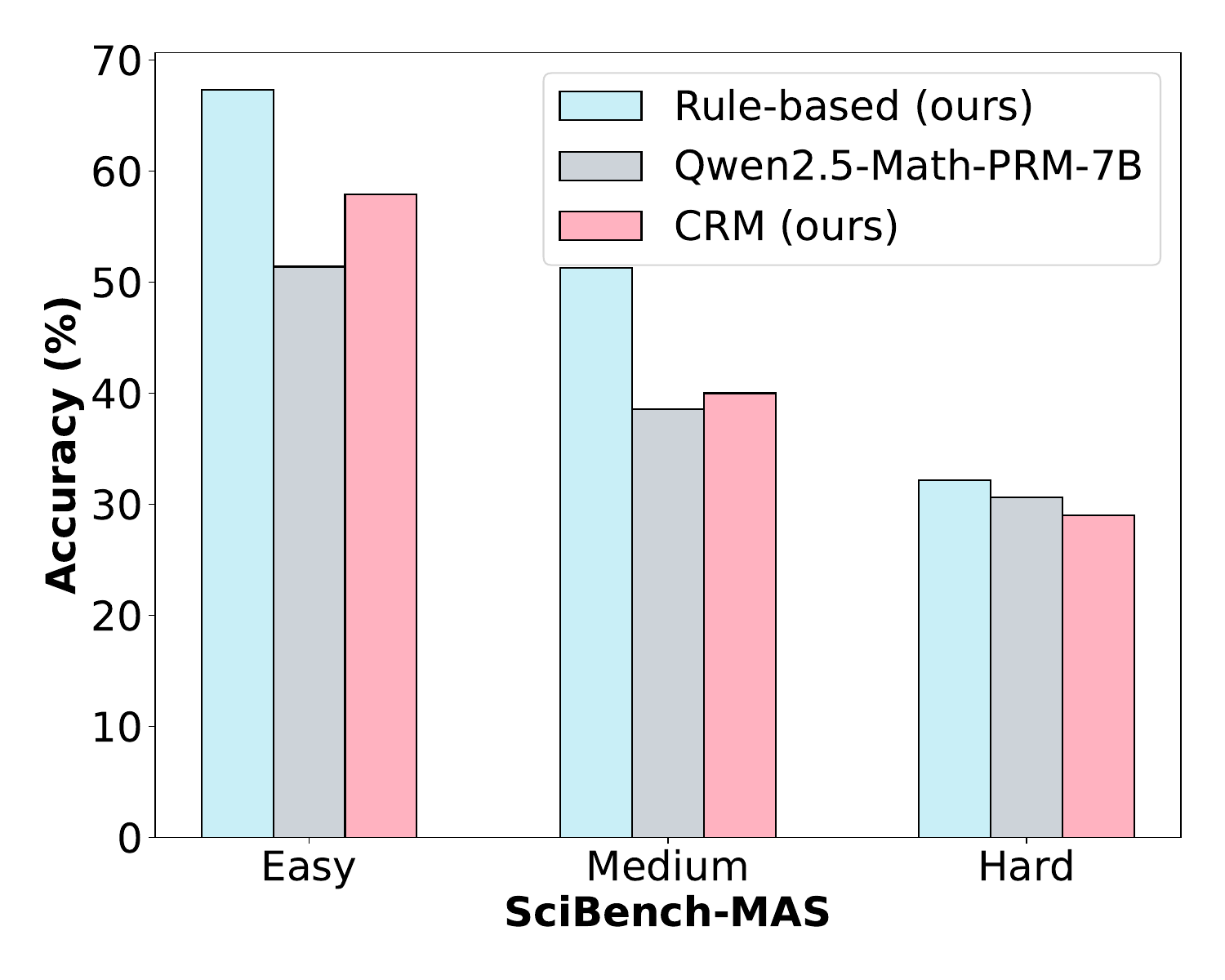}
        \label{fig:ablation_reward}
    }
	\caption{Results of ablation studies. (a) Fine-tuning on domain-specific training data can significantly improve the decomposition quality, thus enhancing overall system performance. (b) Our robust agent selection strategy within the MAS is significant to the performance. (c) Compared to general reward models, our fine-tuned reward model is more task-specific and brings more precise reward signals, thus improving the system performance.}
	\label{fig:ablation_studies}
\end{figure*}

\subsection{Ablation Studies}
\label{subsec:ablation}
We conduct ablation studies on our proposed multi-agent system, examining three core designs: task decomposition, agent selection, and reward signal.

\paragraph{Task Decomposition} We compare three different approaches to task decomposition: (1) \textbf{Ground Truth}, representing an upper bound with human-crafted, meticulously designed task breakdowns; (2) \textbf{GPT-4o}, which autonomously decomposes complex tasks into sub-tasks without targeted fine-tuning; and (3) \textbf{Qwen2.5-7B-SFT}, a model fine-tuned on our dataset based on Qwen2.5-7B, specifically adapted to generate more effective decompositions for complex questions. Figure~\ref{fig:ablation_decomposition} presents the reasoning accuracy under different decomposition strategies. The ground-truth decomposition consistently yields the highest accuracy, underscoring the critical role of precise subproblem segmentation. Meanwhile, the fine-tuned task generator surpasses the naive GPT-4o approach, demonstrating that even a small amount of domain-specific training data can significantly improve decomposition quality and enhance overall system performance.

\paragraph{Agent Selection} We compare three strategies for agent selection: a \textbf{random} strategy, a \textbf{greedy} strategy that always selects the most matching profile, and our proposed \textbf{ReSo} approach. As shown in Figure~\ref{fig:ablation_selection}, \textbf{ReSo} significantly outperforms other strategies across all the datasets, which emphasizes the importance of a robust agent selection strategy within the multi-agent framework. By strategically assigning each sub-task to the most suitable agent, the system can handle increasingly complex tasks with markedly better accuracy.

\paragraph{Reward Signal Ablation}
\label{app:reward_signal}
We investigate the impact of different reward signals on system optimization, considering three approaches. Figure~\ref{fig:ablation_reward} presents the results of training our MAS under these reward schemes on the SciBench-MAS dataset.
Detailed in Appendix \ref{app:RS}
\subsection{Scalability Analysis}

\paragraph{Agent Scalability}
ReSo’s modular design allows the dynamic addition of new agents without retraining the entire system. Each agent registers its static profile in the Dynamic Agent Database (DADB) and begins contributing immediately. For example, during our HumanEval experiments, we simply added some code‐specialist agents on top of the existing 63 agents. ReSo seamlessly leveraged its capabilities to improve overall performance.

\paragraph{Task and Domain Generality}
ReSo is task‐agnostic and domain‐agnostic: as long as domain‐specific data is available, it can generate a task DAG, select appropriate agents, and optimize their collaboration. Our automated data synthesis pipeline converts LLM benchmark into a multi‐step MAS task without human annotations, enabling straightforward migration from mathematics and scientific reasoning to other fields.

\begin{figure}[H]
    \includegraphics[width=0.9\linewidth]{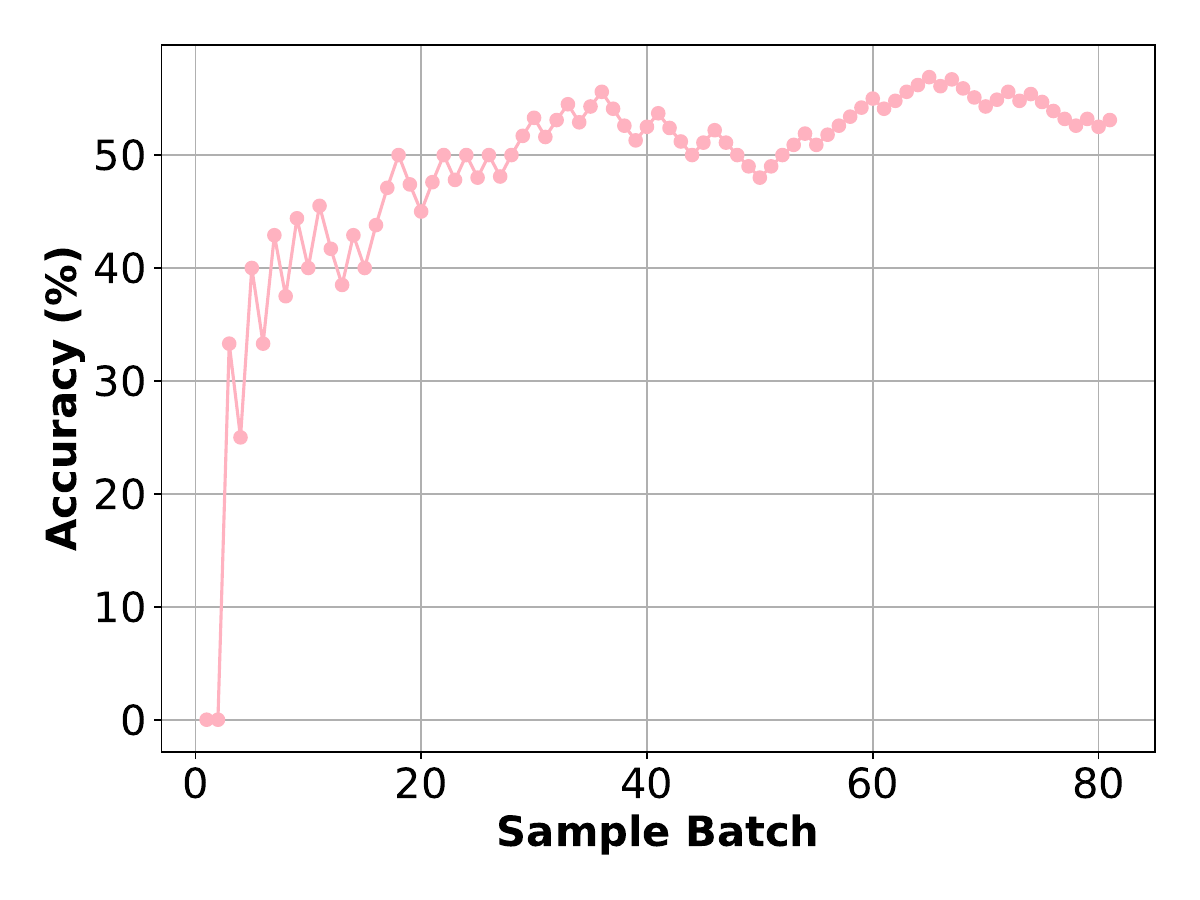}
    \caption{Training Curve of ReSo.}
    \label{fig:training_curve}
\end{figure}

\paragraph{Training Data Scalability}
The effectiveness of agent selection in ReSo grows with more training data. During training, DADB maintains and updates each agent’s reward statistics and cost estimates. As the number of training samples increases, ReSo builds a more accurate model of each agent’s strengths and weaknesses, resulting in progressively better agent assignments and higher overall accuracy. Figure~\ref{fig:training_curve} shows that ReSo's accuracy increases with the training process

\section{Conclusion}
In this work, we introduce ReSo, a reward-driven self-organizing MAS for complex reasoning. By integrating a collaborative reward model, ReSo automates agent selection and collaboration, improving scalability and adaptability. The automated data synthesis framework eliminates manual annotations. Experiments show that ReSo outperforms existing MAS and single LLM baselines. All codes, models, and data have been open-sourced. We expect ReSo to enable co-optimization of MAS and LLM to further enhance reasoning capabilities.
\newpage
\section{Limitations}
Although the base model for the agents is a fixed model, ReSo has demonstrated strong optimizability and scalability as well as good performance. A further interesting research question is: Can the optimization of MAS be performed together with the optimization of a single LLM agent? Specifically, can the reward signal given to the model by our CRM in each step of cooperation be combined with the reinforcement learning-based post-training of a single model to further optimize MAS at both the macro and micro levels? This means a dynamic agent cooperation network, where agents can not only learn how to interact with each other but also fine-tune their weights through feedback from cooperation. We look forward to follow-up research.

\section{Ethical Considerations}
While our proposed ReSo framework focuses on reasoning tasks in the domains of mathematics and science, it has the potential to be applied in other, possibly unethical, contexts. Such misuse could pose significant threats to human society. We strongly urge readers to carefully consider these ethical implications and to adopt a conscientious approach in the development and application of these methods.

\bibliography{custom}

\appendix
\onecolumn
\section{Related work on LLM Reasoning Policy} Reward model is usually combined with different reasoning policies to enhance its effect such as majority voting \cite{wang2023selfconsistencyimproveschainthought}, Chain of Thought (COT) \cite{wei2023chainofthoughtpromptingelicitsreasoning} and Monte Carlo Tree Search (MCTS) \cite{browne2012survey}. OmegaPRM \cite{luo2024improvemathematicalreasoninglanguage} enhances reasoning with a divide-and-conquer MCTS strategy. ReST-MCTS \cite{zhang2024restmctsllmselftrainingprocess} refines reasoning traces using inferred stepwise rewards. RethinkMCTS \cite{li2024rethinkmctsrefiningerroneousthoughts} improves code generation by leveraging execution feedback. In contrast,  Critical Plan Step Learning \cite{wang2024cplcriticalplanstep} employs hierarchical MCTS to generalize across reasoning tasks. Additionally, AlphaMath \cite{chen2024alphamathzeroprocesssupervision} and TS-LLM \cite{feng2024alphazeroliketreesearchguidelarge} enhance reasoning by incorporating a value model and iterative tree search, with TS-LLM further leveraging an AlphaZero-like framework and policy distillation.

\section{Model Performance}
\label{app:single_llm}
\begin{figure}[ht]
    \centering
    \includegraphics[width=0.6\linewidth]{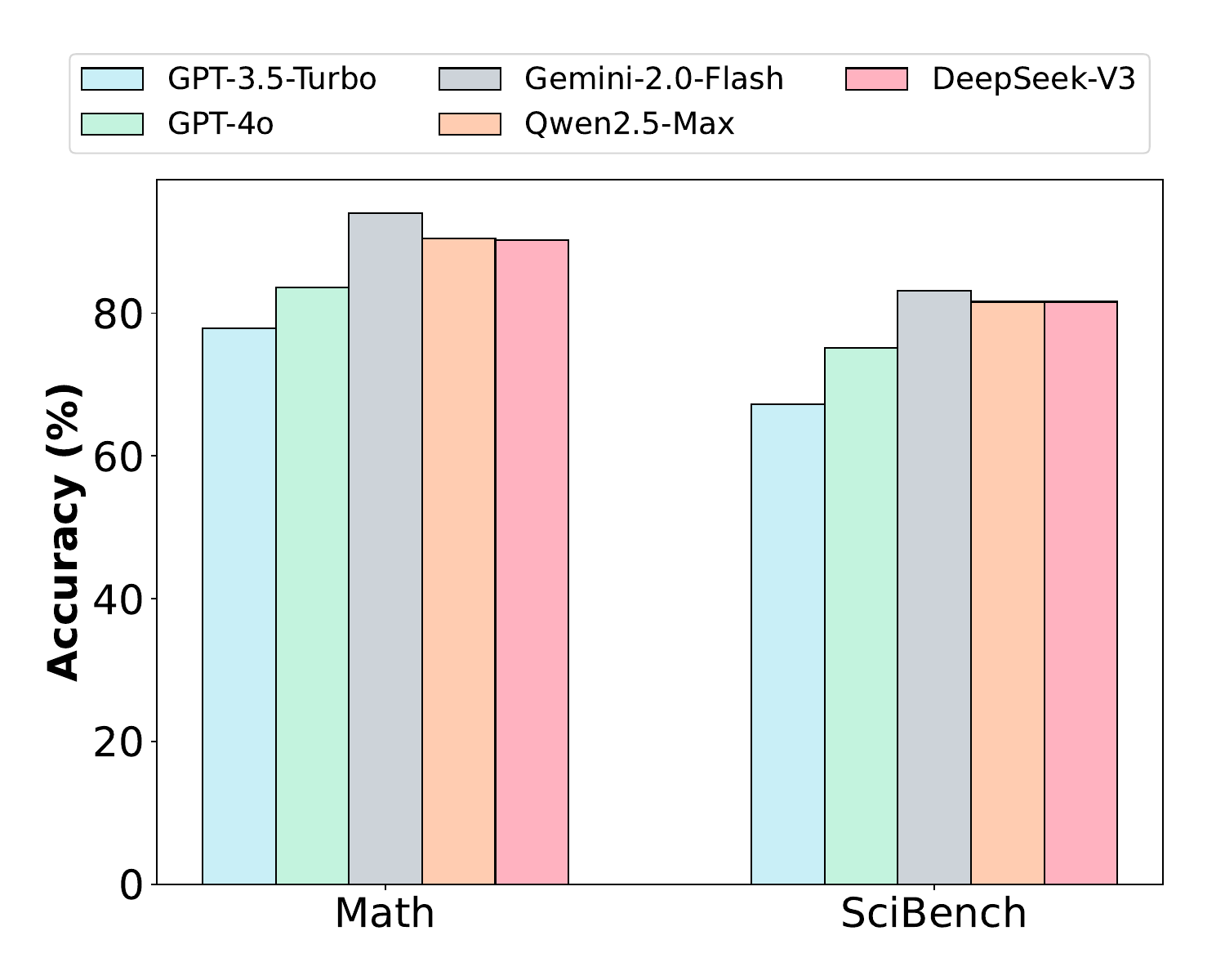}
    \caption{Performance of different models on our selected Math and SciBench dataset subproblems.}
    \label{fig:model_comparison}
\end{figure}

\section{Case Study}

\begin{tcolorbox}[title=Complex Task Synthesis Case Study, breakable, width=\textwidth]
Original Question:

A model for the surface area of a human body is given by
\[
  S = 0.1091\,w^{0.425}\,h^{0.725}.
\]
When ultraviolet radiation of wavelength UNK\_0 (where UNK\_0 = Answer[2] + 56.10 nm) strikes the skin, …; a muscle fiber contracts by 3.5 cm and lifts a weight, assuming Hooke’s law
\(F=-kx\)
with
\(k = \mathrm{UNK\_1} = \text{Answer[0]} + 736.00\);
finally, please calculate
\[
  \text{Answer[0]} \times \text{Answer[1]} \times \text{Answer[2]}
\]
and conclude: “The answer is therefore \(\boxed{[ANSWER]}\).”

\medskip
\textbf{Decomposed Task Graph:}
\begin{itemize}
  \item Task 1 (no deps): Compute \(S\), record as Answer[2].
  \item Task 2 (dep: 1): Set \(\mathrm{UNK\_0} = \text{Answer[2]} + 56.10\), compute UV result, record as Answer[0].
  \item Task 3 (dep: 2): Set \(\mathrm{UNK\_1} = \text{Answer[0]} + 736.00\), compute work via Hooke’s law, record as Answer[1].
  \item Task 4 (deps: 1,2,3): Compute the product Answer[0]·Answer[1]·Answer[2] and box the result.
\end{itemize}

\medskip
\textbf{Agent Routing:}
\begin{itemize}
  \item Task 1 (Calculus)→ \emph{gemini-2.0-flash-exp\_GeometryExpert}
  \item Task 2 (Matter)→ \emph{gpt-4o\_ElectromagnetismExpert}
  \item Task 3 (Thermodynamics)→ \emph{qwen2.5-max\_Thermodynamics\&OpticsExpert}
  \item Task 4 (Aggregation)→ \emph{gemini-2.0-flash-exp\_AlgebraExpert}
\end{itemize}
\end{tcolorbox}

\section{Prompt}
\begin{tcolorbox}[title=Prompt of Agents in the Pool, breakable, width=\textwidth]
\begin{Verbatim}[breaklines=true, breakanywhere=true, fontsize=\footnotesize]
[gpt-4o_1]
model = gpt-4o
role = MechanicsExpert
prompt = You are a highly knowledgeable mechanics expert in a multi-agent system. You are given a sub-task related to classical mechanics, statics, dynamics, kinematics, or fluid mechanics. First, read and understand the previous questions and answers from other agents. Identify the variables that have already been solved and ensure consistency with their results. Then, systematically break down your sub-task, applying relevant physical laws such as Newton’s laws, conservation principles, or motion equations. Justify your reasoning, verify unit consistency, and cross-check with previous agent outputs before providing a well-explained solution.


[gpt-4o_2]
model = gpt-4o
role = ElectromagnetismExpert
prompt = You are an expert in electromagnetism within a multi-agent system. You are assigned a sub-task related to electric fields, magnetic fields, circuit analysis, or electromagnetic waves. First, read and understand the previous questions and answers from other agents, extract solved variables, and ensure logical consistency. Apply fundamental principles such as Maxwell’s equations, Gauss’s law, or Faraday’s law to solve your sub-task systematically. Clearly outline your steps, justify the assumptions, and verify that your solution aligns with previous agents' work. If discrepancies arise, propose possible resolutions.

[gpt-4o_3]
model = gpt-4o
role = Thermodynamics&OpticsExpert
prompt = You are an expert in thermodynamics and optics in a multi-agent system. Your role is to solve a specific sub-task while ensuring coherence with previous agents' results. First, read and understand the previous discussions, extract solved variables, and align your approach with existing solutions. Apply principles such as the first and second laws of thermodynamics, heat transfer models, or optical laws (e.g., Snell’s law, diffraction, and wave optics). Provide a detailed step-by-step solution, justify calculations, and validate numerical consistency with prior agent outputs. If uncertainties arise, suggest possible clarifications.

[gpt-4o_4]
model = gpt-4o
role = InorganicChemistryExpert
prompt = You are an inorganic chemistry expert operating in a multi-agent system. Your sub-task may involve chemical bonding, periodic trends, reaction mechanisms, or coordination chemistry. Carefully review the previous questions and answers, identify already determined variables, and ensure consistency with past calculations. Apply relevant chemical principles to analyze and solve your assigned problem step by step. Provide balanced chemical equations, validate reaction feasibility, and explain your reasoning clearly. If your results depend on prior agents’ outputs, verify their correctness and suggest refinements if necessary.

[gpt-4o_5]
model = gpt-4o
role = OrganicChemistryExpert
prompt = You are an organic chemistry expert in a multi-agent system, responsible for solving a sub-task related to molecular structures, reaction mechanisms, or synthetic pathways. First, review previous discussions, extract key solved variables, and ensure consistency with prior agent responses. Then, apply organic chemistry principles such as resonance effects, nucleophilic-electrophilic interactions, and reaction kinetics to derive a precise solution. Provide clear mechanistic explanations, reaction diagrams if necessary, and cross-check results to maintain logical coherence within the system.
\end{Verbatim}
\captionsetup[figure]{hypcap=false}
\captionof{figure}{The prompt of agents in the pool.}
\label{figure:prompt of agents}
\end{tcolorbox}

\begin{tcolorbox}[title=Prompt of the Task Plan Generator, breakable, width=\textwidth]
\begin{Verbatim}[breaklines=true, breakanywhere=true, fontsize=\footnotesize]
"""
You are an AI assistant specialized in generating structured prompts for domain-specific experts in a multi-agent system. 

**Task:**  
Given a subquestion, analyze its domain, required expertise, and problem complexity. Then, generate a structured prompt that precisely describes the expert’s role in solving the problem. The generated prompt will be used for vector-based similarity matching to select the most appropriate agent from an agent pool.

**Prompt Format:**  
"You are a [Expert Type], highly skilled in [Specific Knowledge Areas]. Your task is to analyze the problem by first reviewing previously solved variables and solutions from other agents in the multi-agent system. Apply domain-specific knowledge to reason rigorously and provide a well-structured, logically sound answer. If calculations are required, show all steps. If problem decomposition is needed, outline a systematic approach. Ensure consistency with previous solutions in the multi-agent system and resolve any discrepancies when necessary. Your role is to assist in solving complex reasoning problems with precision and alignment with the broader system."

**Instructions for Prompt Generation:**
1. **Expert Type Selection**: Identify the most relevant expert type (e.g., MechanicsExpert, AlgebraExpert, ThermodynamicsExpert).
2. **Specific Knowledge Areas**: Define the precise knowledge fields required to solve the problem.
3. **Problem Scope & Complexity**: Determine whether the problem requires deep theoretical knowledge, numerical computation, or practical modeling.

**Output:**  
Provide only the generated prompt without additional explanations."""
\end{Verbatim}
\captionsetup[figure]{hypcap=false}
\captionof{figure}{The prompt of the task plan generator.}
\label{figure:}
\end{tcolorbox}

\section{Agent Selection Visualization}
The agent selection distribution during the testing phase of Scibench-MAS-Easy reveals that Gemini-2.0-Flash-Exp and Qwen2.5-Max were the most frequently selected models after training.
\begin{figure}[H]
    \centering
    \includegraphics[width=1\linewidth]{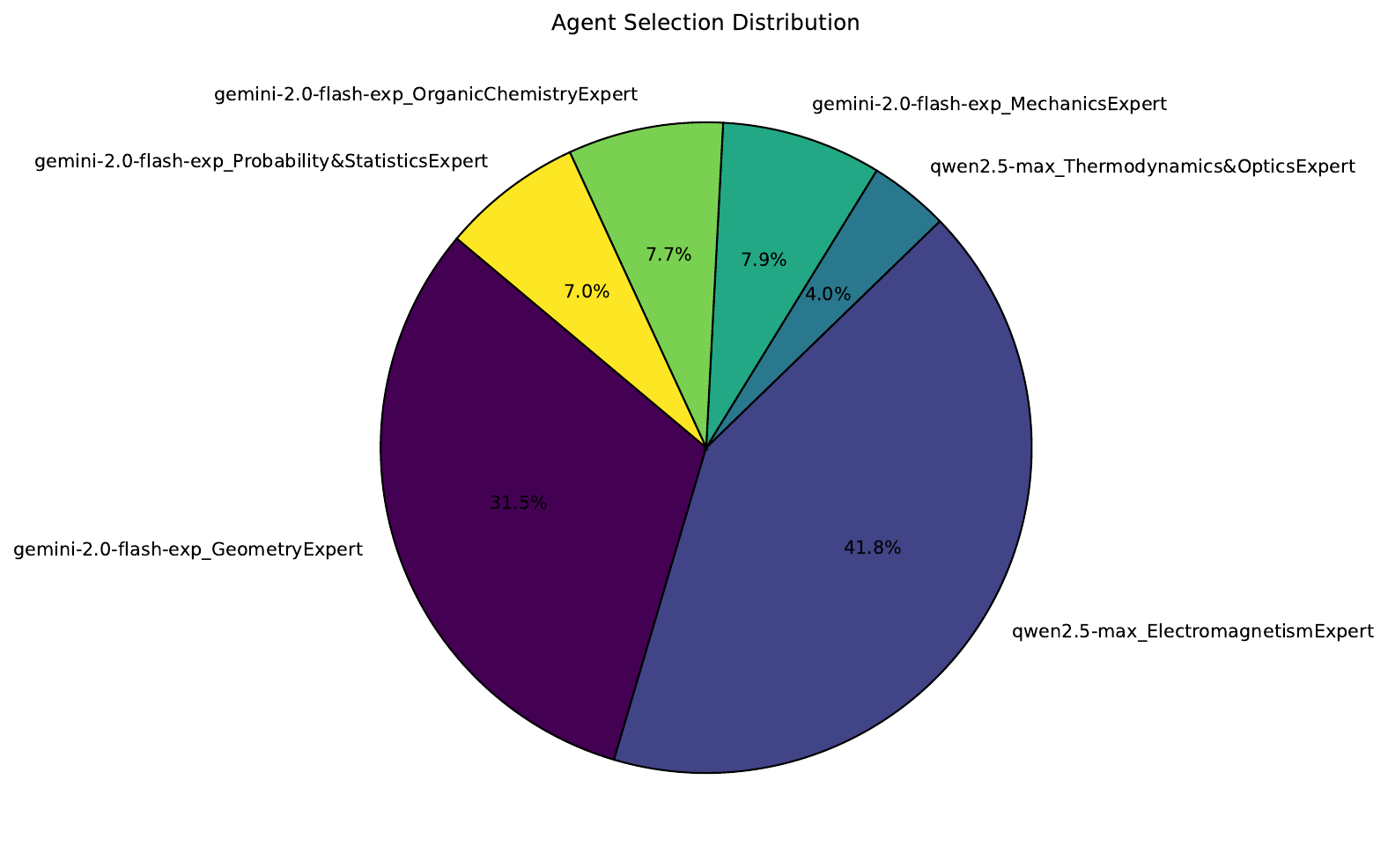}
    \caption{Testing stage on the easy-level tasks in Scibench-MAS.}
    \label{fig:sci3}
\end{figure}

\begin{figure}[H]
    \centering
    \includegraphics[width=1\linewidth]{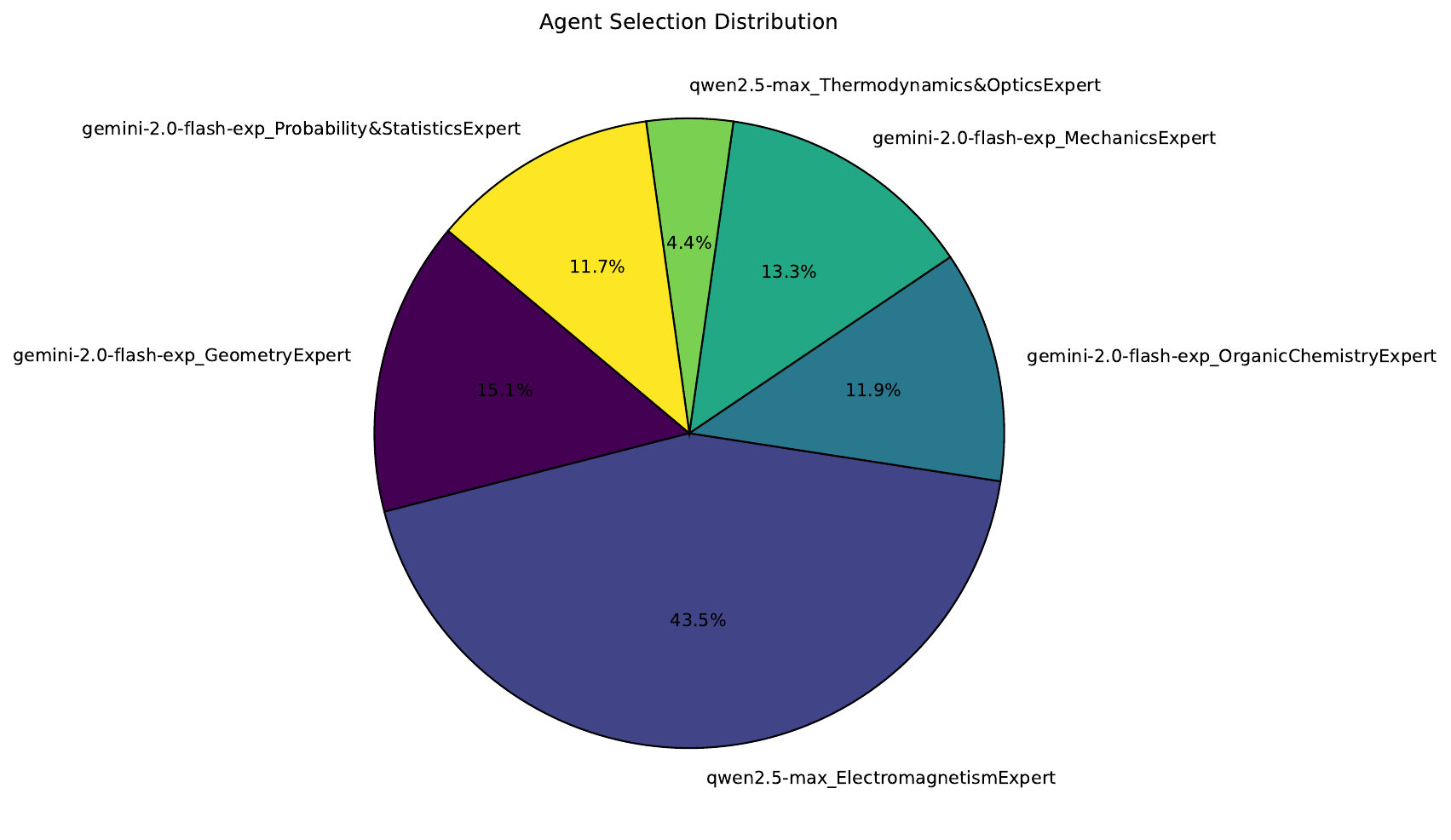}
    \caption{Testing stage on the hard-level tasks in Scibench-MAS.}
    \label{fig:sci7}
\end{figure}

\section{Hyperparameters}
During both training and testing, a set of weighted factors and constraints guide agent selection, allowing for dynamic adjustments. Specifically, \texttt{similarity\_weight} = 0.6 regulates the influence of subproblem-agent similarity, \texttt{reputation\_weight} = 1.0 balances agent selection based on past performance, and \texttt{cost\_weight} = 1.0 accounts for computational overhead. A \texttt{THRESHOLD} = 0.6 establishes the similarity cutoff for specialized handling of certain subproblems, while \texttt{EXPLORATION\_CONST} = 0.3 encourages periodic assignments to underutilized agents. During testing, hyperparameters can be adjusted to fine-tune the selection process—modifying \texttt{similarity\_weight} and \texttt{THRESHOLD} controls the search scope, adjusting \texttt{reputation\_weight} increases the weight of agent reputation in scoring, and tweaking \texttt{cost\_weight} alters the impact of computational overhead, enabling a flexible trade-off between efficiency and performance. Finally, \texttt{TOP\_K} = 3 restricts the number of candidate agents per subproblem, balancing exploration and efficiency in the selection process.

\begin{figure}[H]
    \includegraphics[width=1\linewidth]{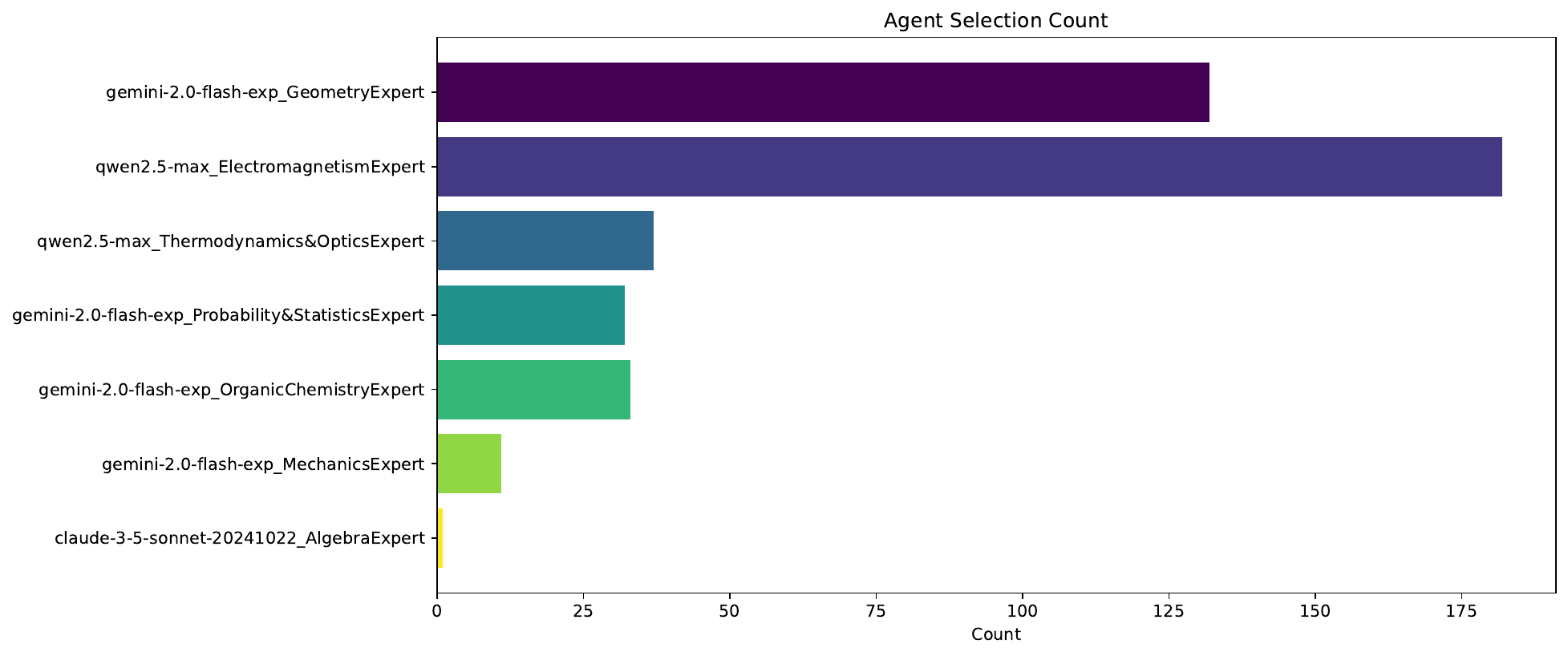}
    \caption{Testing stage on the medium-level tasks in Scibench-MAS using reputation\_weight 1.}
    \label{fig:rb}
\end{figure}

\begin{figure}[H]
    \centering
    \includegraphics[width=1\linewidth]{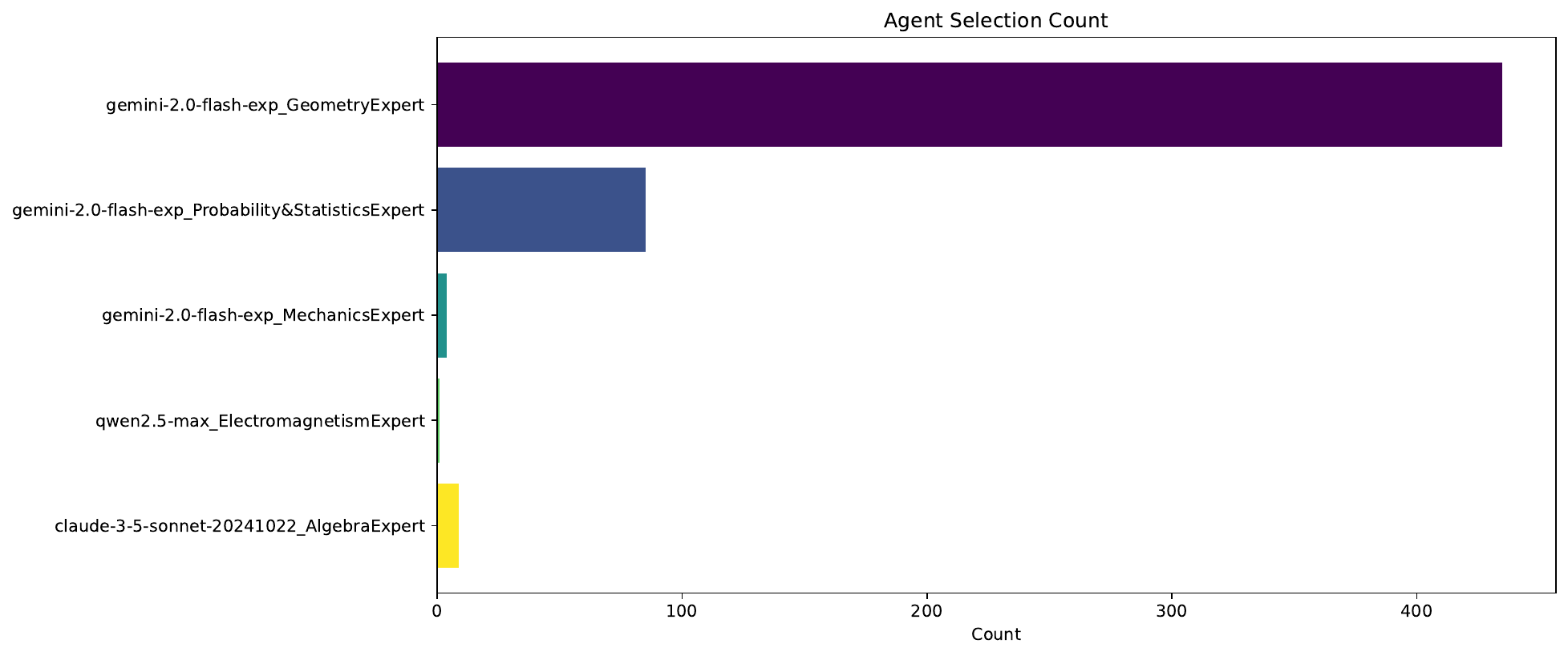}
    \caption{Testing stage on the medium-level tasks in Scibench-MAS using reputation\_weight 2.}
    \label{fig:sci5}
\end{figure}

\begin{figure}[H]
    
    \centering
    \includegraphics[width=0.8\linewidth]{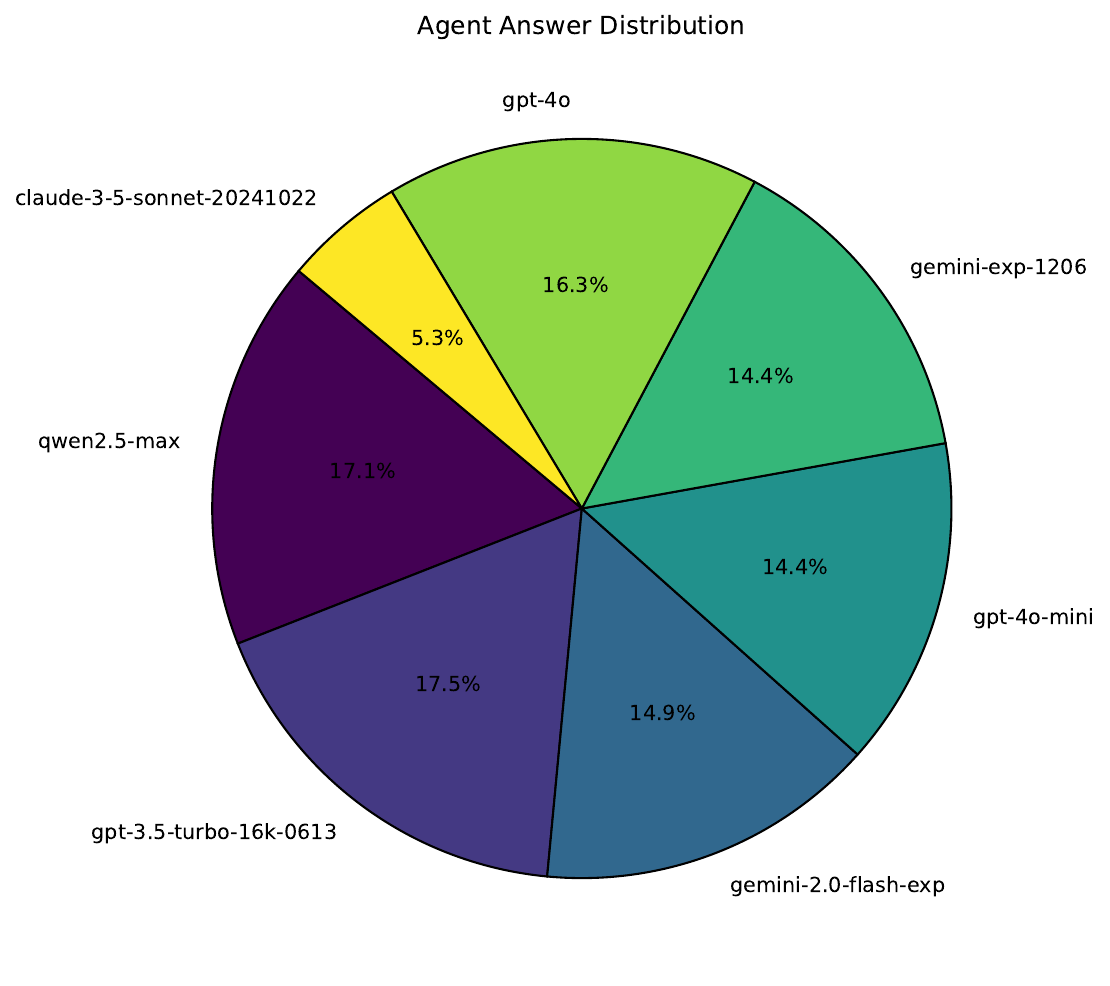}
    \caption{Testing stage on the medium-level tasks in Scibench-MAS without training.}
    \label{fig:random selection}
\end{figure}

\paragraph{Token Efficiency} Table~\ref{tab:model_performance} also compares the average number of tokens consumed per task. ReSo maintains a relatively moderate token usage, which is significantly lower than certain baselines like DyLAN (14.6k vs 64.1k, 20.7k vs 77.8k). This balance between performance and computational cost underlines ReSo's practical efficiency in real-world, large-scale scenarios.

\section{Reward Signal}
\label{app:RS}
We investigate the impact of different reward signals on system optimization, considering three approaches: (1) \textbf{Rule-based}, which provides strictly accurate, predefined evaluations for sub-task solutions; (2) \textbf{General Reward Model}, using Qwen2.5-Math-PRM-7B as a reward function without task-specific fine-tuning; and (3) \textbf{Fine-tuned Reward Model}, i.e., our CRM proposed in ~\ref{sec:crm}. Figure~\ref{fig:ablation_reward} presents the results of training our MAS under these reward schemes on the SciBench-MAS dataset. The rule-based reward yields the best results, confirming the importance of precise reward signals. Besides, our CRM brings a slight improvement compared to the original Qwen2.5-Math-PRM-7B model. We also observe an instance of \emph{reward hacking} when using the Qwen reward model: specifically, Qwen2.5-Max tends to receive inflated scores when acting as the reasoning agent. As a result, during inference, the MAS disproportionately selects Qwen2.5-Max to handle sub-tasks, even in cases where it does not necessarily produce the best solutions.

\section{CRM,ORM,PRM}
\label{app:crm}
Our Cooperative Reward Model (CRM) is inspired by OpenAI’s PRM, but it has been extended and adapted to the multi-agent system (MAS) setting. In our complex tasks, multiple sub-tasks exist, and the CRM scores each sub-task’s response based on the outputs from prior agents. While conceptually similar to PRM—where each sub-task can be seen as a step—PRM cannot be directly applied to our MAS setting due to fundamental structural differences.

\section{Comparison with Chain-of-Thought (CoT) Methods}
We would like to clarify that the prompts used in our single-model evaluation experiments already support step-by-step reasoning, thus reflecting Chain-of-Thought (CoT) style outputs. These models are capable of multi-step reasoning and demonstrate CoT-style thinking when tackling complex problems. However, as demonstrated in our results, these CoT-style single-model approaches perform poorly on tasks with high complexity and combinatorial reasoning. As task difficulty increases, even the strongest single LLMs exhibit a significant drop in accuracy—approaching 0\% at the highest difficulty level. This clearly indicates that "step-by-step thinking" alone is insufficient for solving the kinds of deep combinatorial reasoning tasks we designed. Our proposed method, ReSo, substantially outperforms these CoT-style baselines. In addition, ReSo introduces structural and functional advantages over traditional CoT methods. CoT follows a linear reasoning path, whereas ReSo constructs a task graph composed of multiple subtasks, each solvable independently by different expert agents. This allows for horizontal task expansion and fine-grained skill decomposition. A key limitation of CoT is its dependence on a single model’s context length, reasoning capabilities, and domain knowledge. ReSo addresses these limitations by decomposing tasks, dynamically routing them, assigning subtasks to the most appropriate agents, and using reward mechanisms to drive learning.

\section{Qwen Model Dependence}
We would like to clarify that the performance gains observed in ReSo primarily stem from the task decomposition and multi-agent cooperation architecture, rather than solely from a stronger base model. Our approach consists of two stages. The first stage uses an LLM to decompose the task, and the second stage selects the most suitable agents to handle the subproblems. To further demonstrate the effectiveness of our framework, we conducted a new experiment. Even when Qwen-sfted is used for task decomposition, single-agent approaches still fail. This emphasizes that cooperation among agents is necessary. Additionally, our fine-tuned Qwen-7B model performs comparably to GPT-4o for task decomposition, but it is only when subtasks are assigned to specialized agents that the system achieves significant improvements in performance.
\begin{table}[ht]
  \centering
  \caption{Qwen model dependence}
  \label{tab:qwen-model-dependence}
  \begin{tabular}{lccc}
    \toprule
    model & Easy & Medium & Hard \\
    \midrule
    Qwen-sfted + (no ReSo) single agent & 27.5 & 5.6  & 4.5  \\
    GPT-4o + ReSo                       & 71.4 & 43.8 & 34.8 \\
    Qwen-sfted + ReSo                   & 79.1 & 56.2 & 33.7 \\
    \bottomrule
  \end{tabular}
\end{table}

\section{Computational Complexity and Runtime}

\paragraph{Inference Parallelism.}
Independent DAG subnodes can be executed in parallel, mitigating runtime overhead. Despite a higher token usage, ReSo achieves greater accuracy gains, justifying the cost:

\begin{table}[ht]
  \centering
  \caption{Token usage and runtime comparison}
  \label{tab:runtime-comparison}
  \begin{tabular}{lcc}
    \toprule
    Method    & Tokens & Time (h) \\
    \midrule
    MetaGPT   & 16.1 k & 3.2 \\
    DyLAN     & 64.1 k & 8.0 \\
    GPTSwarm  & 14.9 k & 1.3 \\
    GDesigner & 16.9 k & 4.0 \\
    ReSo      & 25.9 k & 4.1 (3 training + 1.1 testing) \\
    \bottomrule
  \end{tabular}
\end{table}

\end{document}